\documentclass[10pt]{article}
\usepackage{graphics}

\begin{document}   

\title{On the Nature of the Phase Transition Triggered by 
Vortex-Like Deffects in the 2D Ginzburg-Landau Model}

\author{G. Alvarez and H. Fort \\
Instituto de F\'{\i}sica, Facultad de Ciencias, \\ Igu\'a 4225, 11400
Montevideo, Uruguay}

\maketitle

\begin{abstract}

The two dimensional lattice Ginzburg-Landau hamiltonian is simulated 
numerically for different values of the coherence length $\xi$ in units of the 
lattice spacing $a$, a parameter which controls amplitude fluctuations.
The phase diagram on the plane $T-\xi$ is measured. 
Amplitude fluctuations change dramatically the 
nature of the phase transition: for   
values of $\xi/a \simeq 1$
, instead of the smooth Kosterlitz-Thouless
transition there is a {\em first order} transition
with a discontinuity in the vortex density $v$ and a
sharper drop in the helicity modulus $\Gamma$.
Both observables $v$ and $\Gamma$ are
analyzed in detail at the crossover region between first and second 
order which occurs for intermediate values of $\xi/a$.
\end{abstract}

\vspace{1mm}

pacs: 74.20.De, 74.60.-w, 64.60.Cn

\vspace{2mm}
\section{Introduction}
Several condensed matter systems undergo phase transitions which are
triggered by vortex-like defects.
For instance, this is the case of superfluid helium, 
superconductors and the melting transition. 
In particular, the discovery
of high-temperature superconductors has boosted the interest in
the behavior of vortex lines in the mixed state and opened
a new area which regards the physics of vortices as a new state
of matter \cite{bl-fe-ge-la-vi94}. 
The Ginzburg-Landau (G-L) model involving a complex field $\psi =
|\psi|\exp[i\theta]$ captures the essence of all the above condensed 
matter systems.  

A common assumption when studying vortex-like topological excitations 
(singular phase $\theta$ configurations)
is that amplitude fluctuations can be neglected and that the only
relevant degree of freedom is the phase angle $\theta$.  
The ``phase only" approximation is equivalent to consider the limit
in which the quartic coefficient of the G-L hamiltonian $u \rightarrow
\infty$ which freezes the amplitude or equivalently to take the coherence
length $\xi=0$. On the lattice this is the well known XY model.
Thanks to the work of Berezinskii \cite{be72} and Kosterlitz and Thouless 
\cite{ko-th73} in the early 70's we have a fair understanding of the
XY model in two dimensions. It turns out that a topological
phase transition, the so called Kosterlitz-Thouless (KT) transition,
takes place driven by the unbinding of vortex-antivortex pairs 
at a temperature $T_{KT}$. A crucial ingredient of the KT transition
is the logarithmic interaction between vortices.
For that reason in ref. \cite{ko-th73} the authors concluded that in 
a charged superfluid, due to screening, vortices will always unbind at nonzero
temperature. Latter on, it was realized that vortices in sufficiently
thin superconducting films indeed interact logarithmically 
\cite{be-mo-or79}-\cite{he-fi83} 
and it was pointed out that the KT phase transition might also apply 
to thin-film superconductors. 
The 2 dimensional flux line lattice (FLL) melting transition
was also regarded  as based on a KT-type theory \cite{do-hu79}. 
More recently, it was suggested that some features of the
KT theory might be present in under-doped superconducting cuprates 
\cite{nature99}. 

However, the nature of the phase transition triggered by vortices in 2D
systems still remains under discussion. 
By means of Monte Carlo simulations
of the XY model with a modified nearest neighbor interaction it was
shown that, depending on the value of an additional parameter, 
continuous as well first-order transitions take place 
\cite{jo-mi-ny93}-\cite{hi84}. 
The existence of both kinds of phase transitions is 
in accordance with the richer structure of 
the 2D Coulomb gas found by Minnhagen and Wallin \cite{mi-wa87} using 
self-consistent renormalization group equations and with the tendency
towards first-order transition which develops in the case of
a strong disorder coupling constant \cite{ko92}. 
The 2D FLL also still remains 
controversial theoretically as well as experimentally.  
Experimental works and numerical simulations favor a KT-like 
transition in 
some cases and a discontinuous one in others \cite{st88}.
A recent extensive Monte Carlo simulation found a 
first order transition at a temperature close to the
estimated one assuming a KT melting transition \cite{ka-na93}.

As Bormann and Beck \cite{bo-be94} pointed out,
at $T\simeq T_c$ the amplitude fluctuations may affect the critical 
behavior and cannot be neglected. 
Therefore, instead of the XY model -with fixed amplitude fields-
it is worth investigating the effects of vortices when amplitude 
fluctuations are taken into account. Depending on the chosen parameterization,
amplitude fluctuations are controlled
by the quartic coefficient $u$ of the G-L Hamiltonian or by the coherence 
length $\xi$. 
Hence, in this work we simulate the Ginzburg-Landau model (with an 
amplitude and a phase degree of freedom per point)  
on a lattice of spacing $a$ 
and obtain its two-dimensional 
phase diagram in the plane ($T,\xi$).
We will see that the nature of the phase transition of the G-L model 
diverges dramatically from the KT when the parameter $u$ is chosen 
sufficiently small -which in fact is equivalent to take the coherence 
length $\xi \simeq a$- and that this is connected with the appearance of a 
discontinuous jump in the number of vortices.

This article is organized as follows. In section II we define the lattice 
model, the observables to be computed as well as describe the used
Monte Carlo procedure.
In section III we report the main numerical  results. Section IV is devoted to
conclusions and final comments.

\section{Description of the model, the observables and the simulation}

We will consider square lattices of size $L$ and spacing 
$a$ and denote the lattice sites by $x$ and the lattice links 
by ($x,\mu)$ with $\mu=1,2$.
A straightforward discretization of the continuum Landau-Ginzburg 
Hamiltonian produces the expression  \cite{ba80}: 
\begin{equation} 
\beta H^c = \beta a^2 \sum_x [ \sum^2_{\mu=1} \frac{\hbar^2}{2m} (
\psi^c_{x+a\mu}-\psi^c_x)^2/a^2 +r |\psi^c_x|^2 + u |\psi^c_x|^4 ],
\label{eq:H1}
\end{equation}
where  
the superscript $c$ denotes the ordinary parameterization
in the continuum theory, $\beta=\frac {1}{k_B T}$, 
$m$ is the effective mass of the carriers
and the coefficients $r$ and $u$
are analytic functions of the temperature, with $u >0$ for stability. 
Instead of working with this parameterization of
the Hamiltonian -as we did in ref. \cite{AF99}- it is better to
introduce a dimensionless order parameter: $\bar{\psi}_x =
\left(\frac{2u}{|r|}\right)^{\frac 12}\psi^c_x
\equiv \frac{\psi^c_x}{|\psi_{\infty}|}$, where $|\psi_{\infty}|$
is conventionally used because $\psi$ approaches this value
infinitely deep in the interior of the superconductor.
Hence the Hamiltonian can be written as
\begin{equation} 
\beta H^c = \frac {1}{k_B T} H= \frac{1}{2T_L} 
a^2 \sum_x [ \sum^2_{\mu=1}  
(\bar{\psi}_{x+a\mu}-\bar{\psi}_x)^2/a^2 + \frac{1}{2\xi^2}
(1-|\bar{\psi}_x|^2)^2 ],
\label{eq:H2}
\end{equation}
where the lattice temperature $T_L$ and the 
coherence length $\xi$ are given by
\begin{equation}
\frac{1}{T_L}=\frac{\hbar^2|r|}{2m u k_B T}
\;\;\;\;\;\;\;\;\;
\frac{\xi^2}{a^2}=\frac{\hbar^2}{2m |r|};
\label{eq:connec}
\end{equation}
in what follows we will denote the lattice temperature simply 
by $T$, suppressing the subscript $L$.
In the limit of $\xi/a=0$ (or $u=\infty$) the radial degree of freedom is 
frozen and this model -sometimes said to describe
{\em soft spins} with non fixed amplitude- becomes the XY 
model which is said to describe {\em hard spins} with fixed amplitude. 
A more interesting and less well studied 
limit is just the opposite i.e. $\xi/a \sim 1$ (by (\ref{eq:connec})
small values of the $u$
parameter) which corresponds to large amplitude fluctuations.

We have simulated the Hamiltonian (\ref{eq:H2}) using 
a Metropolis Monte Carlo standard algorithm. The calculations were performed
using periodic boundary conditions (PBC).
In order to increase the speed of the simulation we have discretized
the $O(2)$ global symmetry group to a $Z(N)$ and compared the results with
previous runs carried out with the full O(2) group in relatively small 
lattices. For the case of $Z(60)$ we found no appreciable differences.
Lattices with $L=10, 20, 24, 32, 40$ and (in some cases) 64 were used. 
For $L=10$ and 20 we thermalized with, usually, 20,000-40,000 sweeps and
averaged over another 60,000-100,000 sweeps. For $L=40$ and 64  
larger runs were performed, typically 50,000 sweeps were discarded for 
equilibration and averaged over 200,000 sweeps. We also performed
some more extensive runs near $T_c$ for small values of $u$.  
The errors for the measured observables are estimated in a standard
way by dividing measures in bins large enough to regard them as uncorrelated
samples.

The following quantities were measured: 

{\bf i)} {\em The vortex density v}.

\vspace{1mm}

The standard procedure to calculate the vorticity  
on each plaquette is by considering the quantity
\begin{equation}
m=\frac{1}{2\pi}([{\theta}_1 - {\theta}_2 ]_{2\pi} +
[{\theta}_2 - {\theta}_3 ]_{2\pi} +
[{\theta}_3 - {\theta}_4 ]_{2\pi} +
[{\theta}_4 - {\theta}_1 ]_{2\pi}),
\label{eq:m}
\end{equation}
where $[{\alpha}]_{2\pi}$ stands for $\alpha$ modulo 2$\pi$:
$[{\alpha}]_{2\pi} = \alpha + 2{\pi}n$, with $n$ an integer
such that $\alpha + 2{\pi}n
\in (-\pi ,\pi ]$, hence
$m=n_{12}+n_{23}+n_{34}+n_{41}$.
If $m\neq 0$, there exists a {\em vortex} which is assigned to the
object dual to the given plaquette. Hence in the case d = 2, $*m$, the dual
of $m$, is assigned to the center of the original plaquette $p$.
The vortex ``charge" $*m$ can take three values: 0, $\pm 1$ (the value 
$\pm 2$ has a negligible probability).
$v$ defined as:
\begin{equation}
v=\frac{1}{L^2} \sum_{x} |*m_x|,
\label{eq:v}
\end{equation}
serves as a measure of the vortex density.

\vspace{1mm}

{\bf ii)} {\em The energy density} $\varepsilon=<H>/L^2$ 
{\em and the specific heat} $c_V$.

\vspace{1mm}

Both observables were
computed to measure the order of the phase transition.
$c_V$ was computed using several procedures: simply as
the energy variance per site i.e. \mbox{$c_V=(<H^2>-<H>^2)/L^2$,} as the mean
of the variance over $B$ bins per site and as the temperature 
derivative of the energy density $c_V= lim \frac{\Delta \varepsilon}{T}$.

\vspace{1mm}

{\bf iii)} The {\em helicity modulus} $\Gamma$ \cite{fi-ba-ja73}

\vspace{1mm}

$\Gamma$ measures the phase-stiffness. For a spin system with PBC 
the helicity modulus measures the cost in free energy of imposing a ``twist" 
equal to $L\delta$ in the phase between two opposite boundaries of the system.
$\Gamma$ is obtained in general as a second order derivative of 
the free energy with respect to $\delta$ -which can be regarded as a uniform
statistical vector potential- evaluated for $\delta \rightarrow 0$.
In such a way one gets the following expression \cite{eb-st83},
which generalizes the one introduced in ref. \cite{fi-ba-ja73}, to an order
parameter with amplitude as well as phase variations:
\begin{equation}
\Gamma =\frac{1}{N} \{ <\sum_{<ij>}~' \mid \psi_i\mid \mid \psi_j \mid \cos
(\theta_i - \theta_j)> - \frac{1}{T}
<{[}\sum_{<ij>}~'\mid \psi_i \mid \mid \psi_j\mid \sin
(\theta_i - \theta_j){]}^2>\},
\label{eq:helicity}
\end{equation}
where the primes denote that the sums are carried out over links
along one of the 2 directions (x or y).

\begin{figure}[p]
\centering
\scalebox{0.5}[0.5]{
\includegraphics{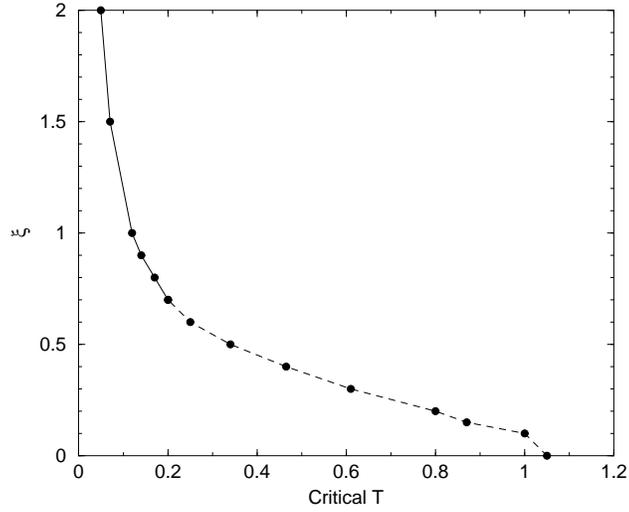}
} 
\caption{Phase diagram in the $(T,\frac{\xi}{a})$ plane
for $L=40$. The phase transition from the ordered
phase to the disordered phase change from second order  
(dashed line) to first order (filled line) at $\xi/a \simeq 0.8$ }
\label{fig:phased}
\end{figure}

\begin{figure}[p]
\centering
\scalebox{0.5}[0.5]{
\includegraphics{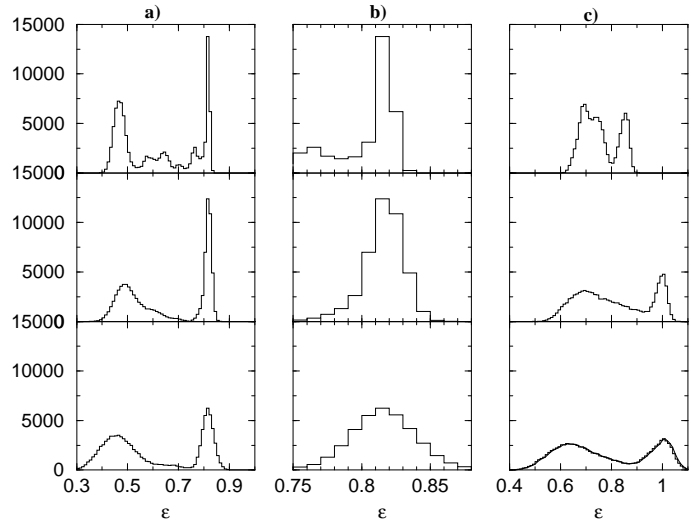}
} 
\caption{Histograms of $\varepsilon$ for $L=10$ (below) ,$L=20$ (middle) 
and $L=40$ (above). a) $\xi/a=0.85$: the 2 peaks becomes sharper
and their position remain fixed as $L$ increases.  
 (b) Zoom of the right peak showing that its width scales as $1/L$. 
 (c) $\xi/a=0.75$: the width of the 2 peaks do not scales as $1/L$
and they approach each other as $L$ increases. }
\label{fig:histo}
\end{figure}

\section{Numerical results}

First, in Fig. \ref{fig:phased} we report the phase 
structure of that model in the
$(T,\xi)$ plane. The intersection point with the temperature 
axis $\xi/a=0$ corresponds to 
the XY standard model.
Around $\xi/a \simeq 0.8$ the phase transition line, separating the
low temperature ordered phase from the high temperature disordered
phase, changes from second order to first order.
In order to illustrate this change in nature we present 
histogram results for two temperatures, one above and one bellow
$T=0.8$ for sizes $L=10,20$ and 40.
The double peak structure of the energy histogram corresponding 
to the 2 coexisting phases, 
characteristic of a first-order transition, is showed in 
Fig. \ref{fig:histo}-a) for
$\xi/a=0.85$. Both peaks remain
fixed as $L$ increases and the width of each
of them clearly scale as $\sqrt(\frac{1}{L^D})=\frac{1}{L}$, 
due to ordinary non-critical fluctuations, as can be checked 
from Fig. \ref{fig:histo}-b). 
On the other hand, for $\xi \le 0.75$ the peaks are much lower
and wider, they move towards to an 
intermediate value of the energy as $L$ increases and their width
do not scale as $\frac{1}{L}$ 
(Fig. \ref{fig:histo}-c)). 

\begin{figure}[p]
\centering
\scalebox{0.5}[0.5]{
\includegraphics{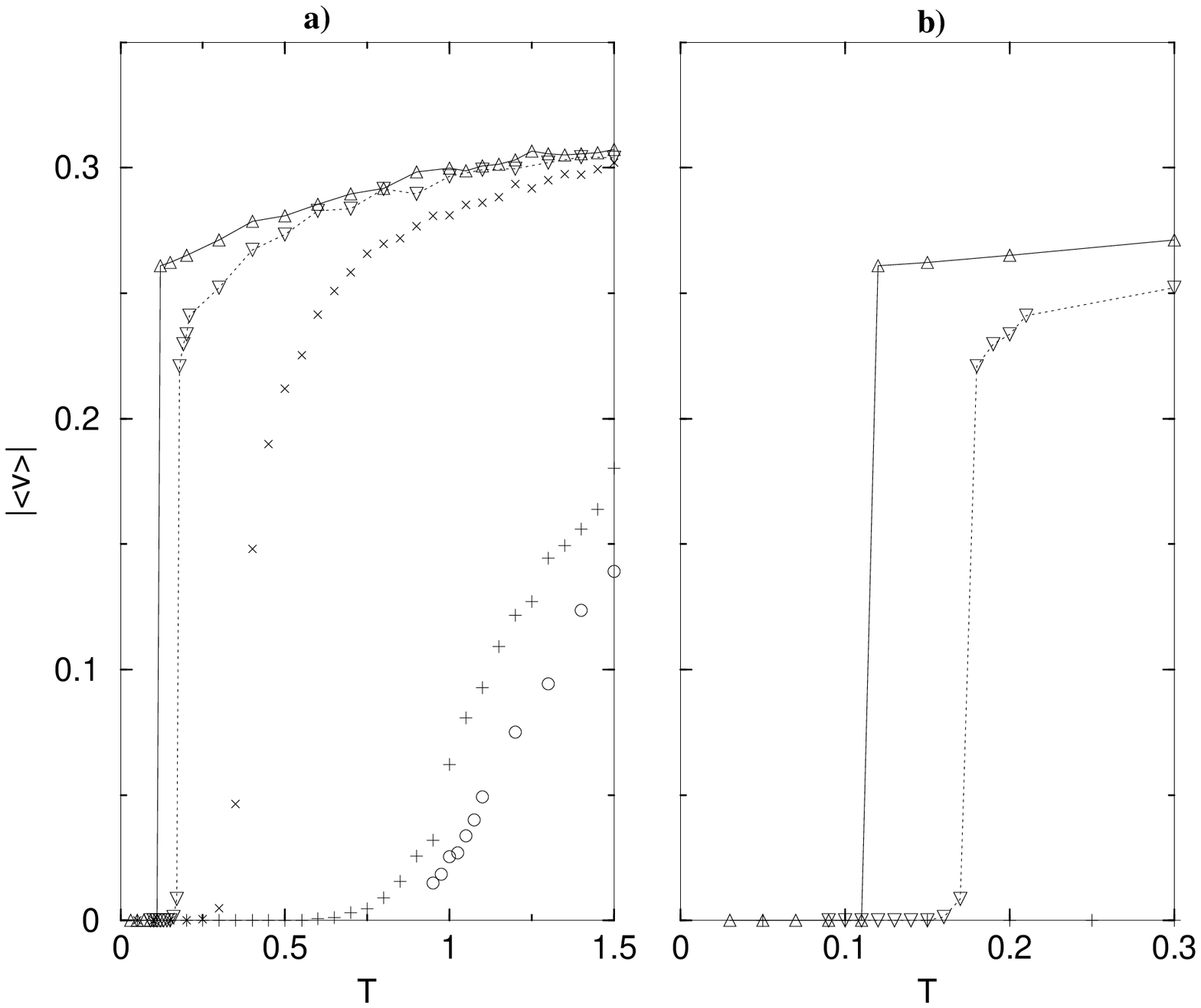}
}
\caption{(a) $v$ vs. $T$ for $\xi/a=1$ (triangles up), 
$\xi/a=0.8$ (triangles down),
$\xi/a=0.5$ ($\times$), $\xi=0.1$ (+), 
and the XY model(circles). (b) Zoom of \ref{fig:allvortices}-(a) 
showing the discontinuity of $v$ at $T_c$ when $\xi/a=1.$}
\label{fig:allvortices}
\end{figure}

The central role played by vortex excitations
in triggering and determining the nature of the phase transition
can be seen in 
Fig. \ref{fig:allvortices}-(a) where $v$ is plotted vs. $T$ for different values of $\xi$
and $L=20$.
For $\xi/a=1$  
we observe a sharp jump in the vortex density $v$
(triangles up). As long as we
decrease $\xi$ the jump becomes more smooth and moves to higher values of 
$T_c$ until for $\xi/a=0.1$ (+ symbols) we get something very close
to the KT behavior (circles).
The increase in the density of vortices when amplitude fluctuations are
large, which is in agreement with the analytical computations 
of ref. \cite{bo-be94}, is due basically to the fact that 
amplitude fluctuations decrease the energy of vortices enhancing vortex production.
The same happens for the XY model with modified interaction \cite{hi84};
in fact, the shape modification can be 
straightforwardly connected to a core energy variation.
It is interesting to note that 
when $\xi\simeq a$ the transition occurs almost from 0 vortex 
(no bound pairs of vortex-antivortex) to a plasma of vortices. 
(see Fig. \ref{fig:allvortices}-b)).
When measuring $v$ very close to $T_c$ from below we found that
for $\xi=1$, $n_{pair}\equiv v/2$ becomes 0 except for a 
very narrow interval -from $T=0.115$ to $T=Tc=0.12$-
in contrast to what happens  
in the case of the K-T transition where bound pairs exist for a 
relatively wide temperature range -from $T=0.7$ to $T=T_c\simeq 1$-.
It is interesting to compare the value of $v$ for 
different values of $\xi$ in all the  
cases for the same very small reduced temperature $t=(T_c-T)/T_=0.01$. 
We found (for different lattice sizes $L$) that $n_{pair}[\xi/a=1]$ 
is one order of magnitude smaller than the 
$n_{pair}[XY]\equiv n_{pair}[\xi/a=0]$ for the K-T transition; 
specifically:
$n_{pair}^{\xi/a=1}(t=0.01)\simeq 0.0046$ compared with 
$n_{pair}^{\xi/a=0}(t=0.01) \simeq 0.05$. 

\begin{figure}[p]
\centering
\scalebox{0.5}[0.5]{
\includegraphics{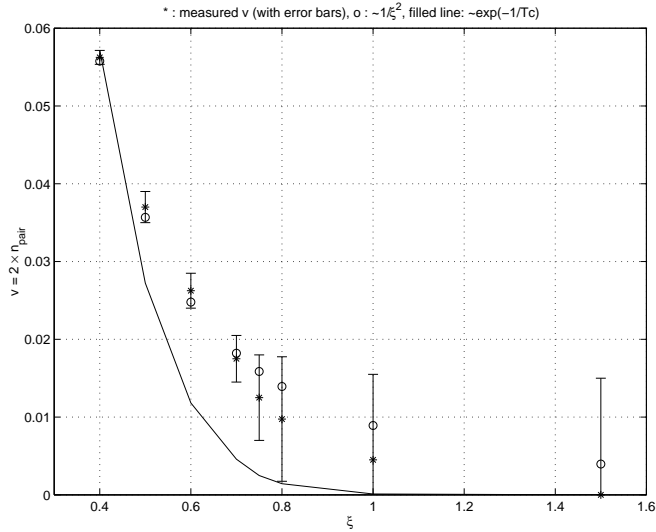}
} 
\caption{$v$ vs $\xi/a$ (asteriscs with error bars) measured on 
a $20 \times 20$ lattice always at the 
reduced temperature $t=0.01$ for that value of $\xi/a$. 
``entropic fitting" $\sim 1/\xi^2$ (circles) and ``energetic fitting"
$\sim\exp[-E_{pair}/T_c]$(solid line).}
\label{fig:vvsxi}
\end{figure}

The scarcity of bound pairs of vortex-antivortex below Tc for the 
large fluctuations regime
($\xi\sim a$) can be explained in terms
of the behavior of their free energy  
$F_{pair}=E_{pair}-TS_{pair}$ at $T \sim T_c$ from below. 
Roughly, $E_{pair} \simeq 2E_c$, where $E_c$ 
is the vortex core energy, and  
$S_{pair} \sim \ln (\frac{L^2}{\xi^2})$. The three quantities
$S_{pair}, E_c$ and $T_c$ all decrease
as $\xi$ increases making difficult to 
disentangle the ``energetic" contribution to $v$, proportional to 
$\exp[-E_{pair}/T_c]$, from the ``entropic" contribution $\sim 1/\xi^2$.
Fig. \ref{fig:vvsxi} shows a plot of $v$ vs. $\xi$ always measured at the corresponding 
reduced temperature $t=0.01$ and fittings with the energetic and entropic behaviors. 
For the intermediate range of $\xi$, although it is not easy to
predict the behavior of the energetic factor, the entropy seems to be
the main responsible for lowering the density of bound 
pairs. 
From $\xi/a \sim 1$, $T_c$ starts to decrease with $\lambda$
faster than $E_{pair}$ and thus 
$\exp[-F_{pair}/T]$ decreases more and more sharply making
smaller and smaller the probability of bound pairs of vortex-antivortex.

\begin{figure}[p]
\centering
\scalebox{0.5}[0.5]{
\includegraphics{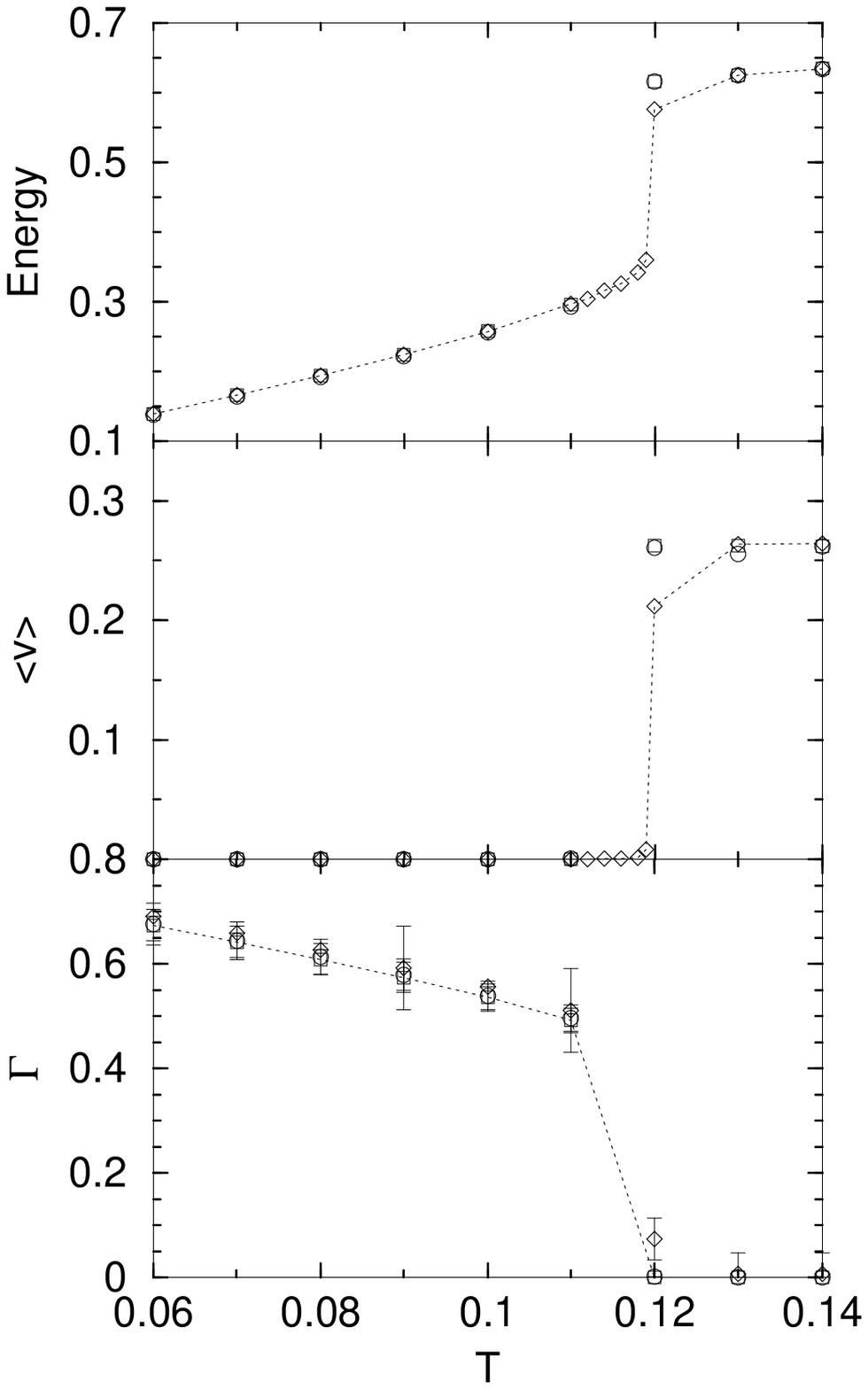}
} 
\caption{$\varepsilon$, $v$ and $\Gamma$ for $\xi/a=1$ for sizes:
$L=10$ (circles), $L=20$ (diamonds) and $L=40$ (squares). Error bars for
$\varepsilon$ and $v$ are smaller than the symbol sizes.}
\label{fig:1}
\end{figure}


\begin{figure}[p]
\centering
\scalebox{0.5}[0.5]{
\includegraphics{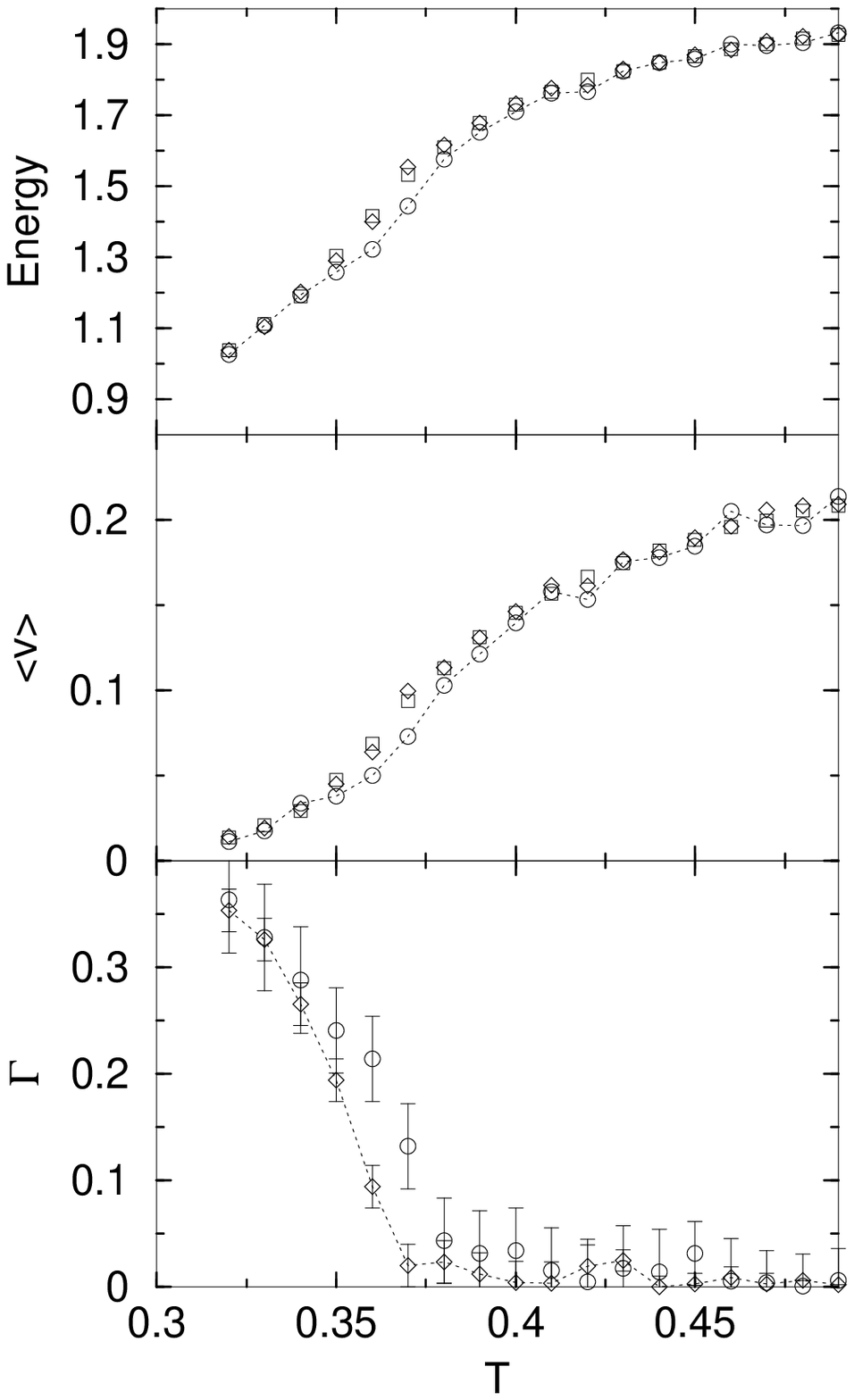} 
}
\caption{$\varepsilon$, $v$ and $\Gamma$  for $\xi/a=0.5$ for sizes:
$L=10$ (circles), $L=20$ (diamonds) and $L=40$ (squares). Error bars for
$\varepsilon$ and $v$ are smaller than the symbol sizes.}
\label{fig:05}
\end{figure}

\begin{figure}[ht]
\centering
\scalebox{0.5}[0.5]{
\includegraphics{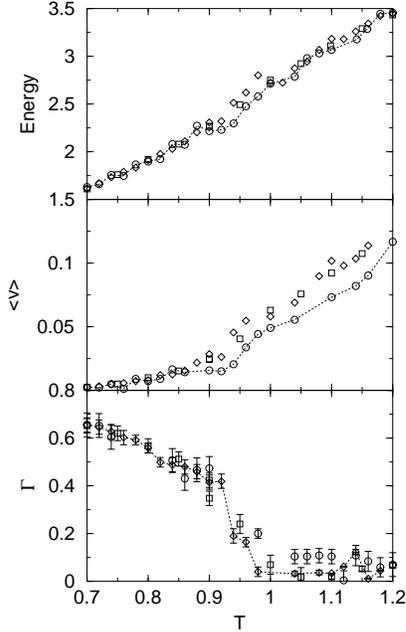}
} 
\caption{ $\varepsilon$, $v$ and $\Gamma$ for $\xi/a=0.1$ for sizes: 
$L=10$ (circles), $L=20$ (diamonds) and $L=40$ (squares). Error bars for
$\varepsilon$ and $v$ are smaller than the symbol sizes.}
\label{fig:01}
\end{figure}


Figures \ref{fig:1}-\ref{fig:01} show plots of $\varepsilon$, $v$ and $\Gamma$ 
for $\xi/a=1, 0.5$ and 0.1. 
For $\xi/a =1$ the transition is clearly first-order:
we observe latent heat and 
discontinuous changes in $v$ and $\Gamma$ at $T=T_c$. 
On the other hand, 
for $\xi/a=0.1$ 
the results are similar to those of the KT transition. In particular,
the energy is very close to the XY energy.

\section{Conclusions}

Therefore, in the G-L model the nature of the phase 
transition depends on the value of the coherence length parameter 
$\xi/a$ (which controls the thermal fluctuations of vortex cores) 
and the following picture emerges:

1) For $\frac{\xi}{a} > 0.8$
the density of vortices experiments a discontinuous jump which 
coincides with a 
first order transition with large latent heat and a sharp
jump in $\Gamma$ all at a $T_c$ which decreases with $\xi$. This 
is the LGW-regime in which a sudden proliferation of
vortices takes place instead of the
unbinding of the much more smooth KT-regime.

2) For $\frac{\xi}{a}\ll 1$ we get basically the XY
model ($\mid \psi \mid=1 $) and the more subtle 
KT transition (with an unobservable essential singularity in the specific
heat at $T_c$ and a much more small non-universal maximum above $T_c$ 
and a universal jump in $\Gamma$).
The number of vortices and the energy evolve smoothly across the transition.

3) For intermediate values of $\xi$  we have an
interpolating regime between LGW and KT.

Whether or not a large enough increment of
$\frac{\xi}{a}$ or $u$ to alter the nature of the 
phase transition driven by vortices can be accomplished 
by varying some thermodynamic parameter, for instance the pressure, 
is something which deserves investigation.

This change in the nature of the phase transition when amplitude
fluctuations of $\psi$ are non negligible ($\xi/a \sim 1$) is 
in agreement with very recent variational computations for the same model
\cite{be-cu99}. 
Furthermore, we found first order transitions for $\xi/a > 0.8$,
an interval which is included into the less restrictive
of ref. \cite{bo-be94} in which for $\xi/a > 0.5$ the authors 
found that the RG trajectories cross the first order line of Minnhagen's
generic phase diagram for the two-dimensional Coulomb gas. 

Finally, our analysis could shed light on the 
nature of the melting transition.  
It is possible that the order of the melting transition in 2D
depends on the particular conditions and details of the studied 
specimen which in turn translate into considerable
different values of the coherence length $\frac{\xi}{a}$.

\vspace{1mm}
Work supported in part by CSIC, Project No.  052 and 
PEDECIBA.

\end{document}